\documentclass[aps,prl,superscriptaddress,showpacs,twocolumn,floatfix]{revtex4}

\usepackage{bm}
\usepackage{graphicx}
\usepackage{epsfig}
\usepackage{color}
\usepackage{amsfonts}

\begin{document}

\newcommand{\qav}[1]{\left\langle #1 \right\rangle}
\newcommand{\myT}{\Gamma}
\newcommand{\rem}[1]{}
\newcommand{\refe}[1]{(\ref{#1})}
\newcommand{\refF}[1]{Fig.~\ref{#1}}
\newcommand{\refE}[1]{Eq.~(\ref{#1})}
\newcommand{\beq}{\begin{equation}}
\newcommand{\eeq}{\end{equation}}
\newcommand{\beqa}{\begin{eqnarray}}
\newcommand{\eeqa}{\end{eqnarray}}
\newcommand{\cg}{\check g}
\newcommand{\inc}{{\rm inc}}
\newcommand{\Pc}{{\cal P}}
\newcommand{\larghezza}{7.8cm}
\newcommand{\cG}{\check{\cal G}}

\title{Distortion blockade in classical nano-electromechanical resonator}

\author{F. Pistolesi}
\affiliation{Laboratoire de Physique et Mod\'elisation des Milieux
    Condens\'es, CNRS-UJF B.P. 166, F-38042 Grenoble, France}

\author{S. Labarthe}
\altaffiliation[present address: ]{Laboratoire des composantes Microsyst\`emes,
CEA-LETI, 38054 Grenoble, France}
\affiliation{Laboratoire de Physique et Mod\'elisation des Milieux
    Condens\'es, CNRS-UJF B.P. 166, F-38042 Grenoble, France}

\date{\today}

\begin{abstract}
We consider a single electron transistor where the central island can oscillate.
It has been shown that for weak coupling of the elastic and electric
degrees of freedom the position of the island fluctuates with a
small variation of the current through the device.
In this paper we consider the strong coupling limit.
We show that the system undergoes a static mechanical instability
that is responsible for the opening of a gap in the current voltage
characteristics even at the degeneracy point.
We provide an analytical description of the transition point.
We also discuss how the mechanical nature of the suppression of the
current can be probed experimentally by a slow modulation of the
gate voltage.
\end{abstract}

\maketitle

Nanoelectromechanics constitutes a rapidly developing and promising
field of mesoscopic physics
\cite{roukes:2001,cleland:1998,blencowe:2004}.
A particularly important and investigated device is the single
electron transistor with mobile parts
\cite{park:2000,sazanova:2004,sapmaz:2005}.
Due to the Coulomb blockade the interplay between the electrical and mechanical
degrees of freedom influences the current in a sizable way.
When the oscillation amplitude is sufficiently large with respect to
the tunnelling length, the modulation of the tunnelling rates leads to
the shuttle phenomena, where the oscillations becomes synchronized
with the electron tunnelling
\cite{gorelik:1998,novotny:2004,pistolesiFCS:2004,pistolesi:2005}.
If the oscillation does not modify the distance between source and
drain, it may nevertheless induce a modulation of the tunnelling
rates through the variation of the electric potential energy in
space
\cite{chtchelkatchev:2005,armour:2004,blanter:2004,usmani:2006,doiron:2006}.
Typically this effect is due to the position dependence of the gate
capacitance.
If the Coulomb force generated by the variation of the number of electrons by
one is $F$, the distance of the two equilibrium positions of the
island will be $X_o=F/k$ where $k$ is the effective elastic constant
of the island.
The corresponding variation of the elastic energy is $E_E=F^2/k$.
For source-drain bias voltages $V$ of the order of $E_E/e$ (with $e$
the electron charge), the interplay between the electrical and
mechanical degrees of freedom becomes important.
Typically  $E_E$ is very small, but within the range of observation.
For a nanotube of 500 nm length, 1 nm radius, suspended at 100 nm from a gate
$E_E$ is of the order of 100 $\mu eV$.
Previous analytic work concentrate on the case $eV\gg E_E$
\cite{armour:2004,armourNOISE:2004}.
Very recently the current for $eV \geq E_E$ at the degeneracy point
has been considered numerically \cite{doiron:2006}.
In this paper we present an analytic theory that gives a good
description of the whole region $eV\geq E_E$ and arbitrary gate voltage.
We find that sweeping the gate voltage for $eV>E_E$ the current
jumps from 0 in the blocked regions to a finite value in the
conducting regions.
We also discuss the behavior of the system for $eV<E_E$.
We argue how a static mechanical instability blocks the current for
any value of the gate voltage.

%
%
\begin{figure}
    \centerline{
    \includegraphics[height=6cm,angle=-90]{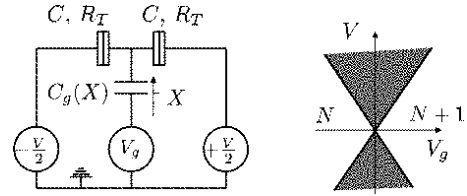}
    }
    \caption{Left panel: Circuit diagram for the single electron transistor with
    an oscillating central island. The variable $X$ parameterizes an effective
    position. Right panel: conducting regions shaded in the $V_g$-$V$ near a degeneracy point
    between $N$ and $N+1$ electrons.}
\label{fig1}
\end{figure}
%
%

Let us consider a single electron transistor with a mobile central island (see \refF{fig1}).
The model has been discussed in details in previous works
(see for instance Refs. \cite{armour:2004,blanter:2004,usmani:2006,doiron:2006}), we discuss the
simplest configuration of symmetric capacitances, resistances, and bias at zero temperature.
(Strictly speaking we assume $E_E \gg  k_B T \gg \hbar \omega $.)
For small oscillations gate capacitance varies linearly in the
displacement of the island $X$.
Charge transport can be described as usual by a master equation where the rates are modulated by
the position of the island.
For positive source drain voltage ($V>0$), and the gate voltage $V_g$
tuned close the degeneracy point between $N$ and $N+1$ electrons on the island,
the non vanishing rates read:
\beq
    \Gamma_{L,R} = f(eV/2 \pm \beta_g(0) eV_g \mp (2N+1)E_C(0) \mp F X)
        \, ,
    \label{eq:gammas}
\eeq
where $\Gamma_L$ and $\Gamma_R$ are the rates for transfer of electrons from the left lead to the
island and from the island to the right lead, respectively,
$R_T$ is the tunnelling resistance of the left and right contacts,
$E_C(X)=e^2/2C_\Sigma$ is the Coulomb
energy of the device as a function of the position of the island
$X$, $C$ and $C_g(X)$ are the junction and gate capacitances, respectively,
$C_\Sigma(X)=2C+C_g(X)$, $f(E)= E/(e^2 R_T)$ for $E>0$ and $0$ for $E<0$,
and finally $\beta_g(X)=C_g(X)/C_\Sigma(X)$.
Choosing $X=0$ as the equilibrium position for the island with $N$ electrons, the force acting
on the island  when an additional electron is added is then
$F = -(2N+1)E_C[1+2C/C_g(0)]C_\Sigma^{-1} dC_g(X)/dX$,
where we set $eV_g=(2N+1)E_C/\beta_g$, since we will consider
small variation of $V_g$ around this value.
The motion of the island is described by the Newton equation:
\beq
    m \ddot X(t)=-k X(t) -F n(t)
    \label{eq:Newton}
\eeq
where $n(t)$ fluctuates stochastically between 0 and 1 according to
the rates given in \refE{eq:gammas}, $m$ is the effective mass of
the island.
It has been shown that the coupling to the electronic degrees of
freedom introduces an intrinsic damping
coefficient \cite{armour:2004}.
We thus do not introduce an extrinsic dissipation into \refE{eq:Newton},
assuming that the intrinsic dissipation is dominant.
It is convenient to introduce reduced variables: $x(\tau)=X/X_o$,
$u=\dot X / (\omega_o X_o)$, $v=eV/E_E$, $v_g= (\beta_g eV_g-(2N+1)E_C)/E_E$,
$\tau= \omega_o t$, $\Gamma_o= E_E/(e^2 R_T \omega_o)$, with
$\omega_o^2=k/m$.
For fixed $x$ the stationary current is given by the usual
expression
\beq
    {I \over R_T V}
    =
    {e\over R_T V} {\Gamma_L \Gamma_R \over (\Gamma_L+\Gamma_R)}
    = {v^2/4-(v_g-x)^2 \over v^2}
    \quad.
    \label{eq:current}
\eeq
In the plane $v$-$v_g$ it leads to the Coulomb diamond structure of
the current, as shown in \refF{fig1}, with the degeneracy point
sitting at $v_g=0$ for $x=0$.

The effect of the oscillation of the central island on the current
can be studied analytically far from the boundaries of the
conducting regions \cite{armour:2004}.
In that limit it has been shown that the probability distribution of
$x$ and $u$ becomes gaussian with a width controlled by the voltage
bias $v$, and the variation of the current due to the mechanical
coupling is always a small part of the unperturbed value.
In this paper we will focus on the regions near the degeneracy point
and near the two lines of transition from the conducting to the non
conducting region (see \refF{fig1} right panel).
We assume that $\Gamma_o \gg 1$ so that the problem can be tackled by
exploiting the separation of time scales between the slow mechanical
oscillations and the frequent electronic hopping.

\begin{figure}
    \centerline{
    \includegraphics[height=7cm,angle=-90]{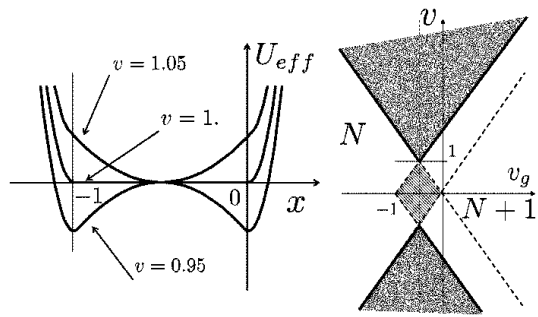}
    }
    \caption{Left panel:
    Effective potential $U_{eff}$ for $v_g=-1/2$ and $v=1.05$, 1, and $0.95$.
    Right panel: Shape of the conducting regions (grey) in the plane $v_g$-$v$
    when the displacement of the mobile part is taken into account.
    The small dashed diamond indicates the mechanically bistable
    region.
    }
\label{fig2}
\end{figure}

We begin with a very simple description of the stochastic force by
substituting into \refE{eq:Newton} the average occupation number:
$n(t) \rightarrow \bar n(x)$, where:
\beq
    \bar n(x) = {v/2+v_g -x \over v}\,, \quad{\rm for}\,|v_g-x|<v/2\,,
    \label{eq:force}
\eeq $\bar n(x)=0$ for $v_g-x < -v/2$,  and $\bar n(x)=1$ for $v_g-x
> v/2$.
This induces an average force on the island that depends on the
occupation of the island itself.
It is convenient to introduce an effective potential to describe
this force (in units of $E_C$):
\beq
    U_{eff}(x) = \int_{x_m}^{x} [x+\bar n (x)]dx
\eeq
where $x_m=(v/2+v_g)/(1-v)$.
The form of $\bar n(x)$ given above
implies that $d^2U_{eff}/dx^2$ equals $(1-1/v)$ for $|v_g-x|<v/2$
and $1$ otherwise.
For $v>1$ the potential has thus a single minimum at $x_m$ with
$-1\leq x_m \leq 0$ and (by construction) $U_{eff}(x_m)=0$.
For $0<v<1$ instead two minima may be present, depending on the
value of $v_g$.
The evolution of the potential as a function of $v$ is shown in
\refF{fig2} for $v_g=-1/2$.
At $v=1$ the potential becomes flat and for $v<1$ it develops two
side minima.
%

%
%
%
%

The form of the effective potential has important consequences on
the current flowing through the device.
Within this simple approximation the average position $x$ is
determined by the minimum of the potential.
The current is then given by \refE{eq:current} with $x=x_m$.
Note that $x_m\leq 0$, this implies that the displacement increases
the gate voltage seen by electrons [cf. \refE{eq:gammas}].
Thus if for $v>1$ we sweep $v_g$ from negative to positive values,
$\bar n=0$ for $v_g<-v/2$, $\bar n$ then starts to increase linearly from
$0$ to $1$.
Due to the displacement of the average position $\bar n$ equals
$1/2$ (degeneracy point) at $v_g=-1/2$ instead of $v_g=0$.
For the same reason it becomes $1$ at $v_g=v/2-1$ instead of
$v_g=v/2$.
The shape of the conducting region changes thus slightly for
$v\gg1$, as shown in \refF{fig2}.
But for $v \leq 1$ the consequence is more dramatic.
The evolution in $v_g$ is no more smooth, due to the presence of two
minima, the system is hysteretic and the stable position and
occupation jump from $x_m=-1$ and $\bar n=0$ to $x_m=0$ and $\bar
n=1$, or viceversa.
The current is thus always blocked for $v<1$.
As shown in \refF{fig2} right panel, a gap appears in the
$v_g$-$v$ plane at the degeneracy point.
This effect can be regarded as the classical counterpart of the
Frank-Condon effect in the quantum limit that has been shown to lead
to a current suppression in molecular devices
\cite{flensberg:2003,weig:2004,koch:2005,mozyrsky:2006}.
The fact that this phenomenon exists also in the classical limit,
makes it probably more easy to be observed in finite temperature
device.
Moreover, the simplicity of the classical description affords an
analytical solution and the the possibility
of treating more complex situation, as the non stationary evolution
discussed in the following.

%
%

\begin{figure}
    \centerline{
    \includegraphics[height=6cm,angle=-90]{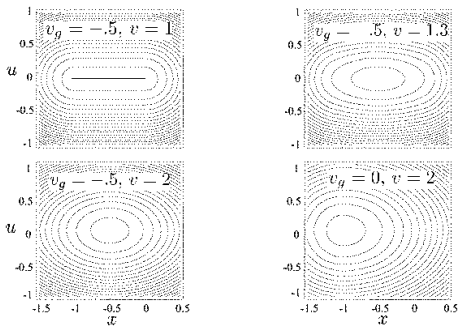}
    }
    \caption{Lines of equal energy and thus of equal probability ${\cal Q}$
    in the $x$-$u$ plane for the values of $v$ and $v_g$ indicated.
    }
\label{fig4}
\end{figure}

Let us now discuss the behavior of the system in the conducting
region ($v>1$).
The dependence of the current and the probability distribution for
$x$ along the degeneracy line ($v_g=-1/2$, and $v>1$) have been
considered very recently in Ref. \cite{doiron:2006} numerically.
We present here an analytical theory that gives a good description
of the current in the whole conducting region, including near the boundaries
and at the apex.
For $\Gamma_o\gg 1$, the electrons have the time to hop in and
out of the grain many times before it can move of a sizable distance.
One can then write a Fokker-Plank equation for the probability
${\cal Q}(x,u)$ \cite{blanter:2004}:
\beq
    {\partial {\cal Q}\over \partial t}
    =
    -{\partial\over\partial u}[F_e(x)-\eta_i(x)u]{\cal Q}
    -{\partial\over\partial x}u{\cal Q}
    +{1\over 2} {\partial \over \partial u^2} S(x) {\cal Q}
    \label{Fokker1}
\eeq
where $F_e(x)=-dU_{eff}/dx$, and $\eta_i(x)={(\partial \bar
n/\partial v_g})/v\Gamma_o$ is the intrinsic damping
\cite{armour:2004,usmani:2006}.
The non vanishing second moment is $S(x)= \int d \tau \qav{\delta
n(\tau)\delta n(\tau)}=2\bar n(x)(1-\bar n(x))/\Gamma_o v$.
It is convenient to simplify further the equation by exploiting the fact that
the fluctuating and dissipative part is small for $\Gamma_o \gg 1$.
The distribution function is thus a function of the effective energy:
\beq
    E_e(x,u) = U_{eff}(x)+u^2/2
    \label{eq:energy}
    \,.
\eeq
From \refE{Fokker1} we can derive an equation for ${\cal
P}(E)\equiv\int dx du {\cal Q}(x,u) \delta(E-E_e(x,u))$:
\beq
    {\partial {\cal P}\over \partial t}
    =
    {\partial \over \partial E}
    \left[
    -\alpha(E) {\cal P}(E)
    +{\partial \over \partial E} \,\left(\beta(E) {\cal P}(E)\right)
    \right]
    \label{Fokker2}
\eeq with $\alpha(E)=\qav{S(x)/2-\eta_i(x)u^2}_{E}$ and
$\beta(E)=\qav{S(x)u^2}_E/2$, where the averages are taken on the
trajectories in the $x$-$u$ plane at fixed energy $E$.
These trajectories are shown for some values of $v$ and $v_g$
in \refF{fig4}.
Near the apex and the borders the shape is quite different from
an ellipse, which is the harmonic oscillator trajectory.
%

%
%
%
%

%
%
\begin{figure}
    \centerline{
   \includegraphics[height=7cm,angle=-90]{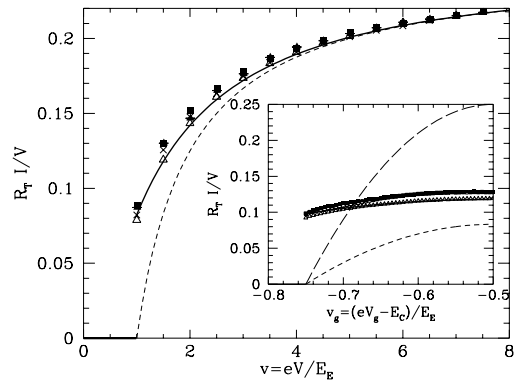}
    }
    \caption{Current as a function of $v$ at $v_g=-1/2$ from the analytical calculation
    (full line), the weak coupling theory (dashed line) and the Monte Carlo simulation
    (points). The values of $\Gamma_o$ are $10$ (square), 1 (triangle), $0.1$ (cross) and $0.01$ (full square).
    In the inset: current as a function of $v_g$ for $v=1.5$. The
    long dashed line is \refE{eq:current} for $x=x_m$.
    } \label{fig5}
\end{figure}
%
%

%
The stationary solution of \refE{Fokker2} is
\beq
    {\cal P}(E) = {\cal N} \exp\left\{ \int^E dE' \alpha(E')/\beta(E')\right\}/\beta(E)
    \label{eq:PE}
\eeq
where ${\cal N}$ is a normalization constant.
Once ${\cal P}(E)$ is known, all physical quantities can be
calculated by averaging first over the trajectories, and then
over the energy.
Let us test the theory at the strongest value of the interaction, at the apex
of the new Coulomb diamond: $v=1$ and $v_g=-1/2$.
The orbits are quite simple for this case (see \refF{fig4}).
The velocity is constant for $|x+1/2|<1/2$ and the trajectory is
exactly half an ellipse for $|x+1/2|>1/2$.
The averages of $\eta_i$ and $S$ on this trajectories lead to the
following expressions for the probability
\beq{\cal P}(E)
    = {\cal N}(1+\pi \sqrt{2E}) e^{-6 E}/\sqrt{E} \,.
    \label{distr}
\eeq
%
The current at the apex can be obtained by integrating
expression \refe{eq:current} with the distribution \refe{distr}.
We find that
$R_T I(v=1,v_g=-1/2)/V=1/(6+2\sqrt{3\pi})\approx 0.0837$.
With the same technique one can calculate the rms of $x$:
$
    \sqrt{\qav{(x-\bar x)^2}} =
    (40 + 3\,{\sqrt{6}}\,{\pi }^{3/2})^{1/2}/
  (96 + 8\,{\sqrt{6}}\,{\pi }^{3/2})^{1/2}
    \approx 0.628.
$

The method can be applied to describe the transition from the
blocked to the conducting state at finite $v$.
In order to keep the calculation simple we consider $v \gg 1$.
For $v_g=-v/2$ the minimum of the potential is at $x=0$.
The potential is harmonic for $x<0$ with oscillation period $2\pi$
and harmonic for $x>0$ with period $2\pi(1-1/v)^{-1/2}$.
The oscillations will thus be more elongated inside the conducting regions (see \refF{fig4}).
The calculation at leading order in $1/v$ gives the distribution:
${\cal P}(E)= {\cal N} [ 1-3 \pi \sqrt{E}/(8 v
\sqrt{2}(1-1/v)^{1/2}) ]^{4(1-1/v)v}$.
With this distribution we find that the current jumps at $v_g=-v/2$
(and symmetrically at $v_g=-v/2-1$) with $R_T I(v,v_g=-v/2^+)/V=
(E_E/eV)\, 8/(3\pi^2)[1-1/2v-1/4v^2+o(1/v^3)]$.
For arbitrary values of $v_g$ and $v$ we found the analytical
expressions for $\alpha$ and $\beta$ and they have been used to obtain
the current by numerical integration.
To check these results we have performed a Monte Carlo simulation of
the model (see for instance Refs.
\cite{pistolesi:2005,doiron:2006}).
\refF{fig5} shows the comparison for $v_g=-1/2$ and $v$ varied
between 0 and 8, and for fixed $v=1.5$ and $v_g$ varied between the
blocked region to the degeneracy point (inset).
We considered values $\Gamma_o$ ranging from $10^{-2}$ to  $10$.
Surprisingly, the agreement between the analytical and the numerical calculations
is very good also for $\Gamma_o \ll 1$.
Apparently the current depends very weakly on $\Gamma_o$ (apart a trivial
linear scaling).

Concerning the $v_g$ dependence of the current (inset of \refF{fig5})
compare the exact result to the weak coupling result of Ref.
\cite{armour:2004} extrapolated to strong coupling (dashed line) or
to the expression \refe{eq:current} taken at $x=x_m$ (long dashed
line).
For $v_g>-v/2$ the current flows through the device inducing a
fluctuation of the position.
The observed current is then the average of \refE{eq:current} over
the values of $x$ visited by the island.
This average increases the current near the threshold, producing the
discontinuity, and reduces it near the degeneracy point.
Also for the rms of the position the analytical estimate
given above agrees with the numerical simulation:
for $\Gamma_o$ varying in the same range of \refF{fig5} we find
that $\sqrt{\qav{x-\bar x}}$ ranges between $0.59$ and $0.64$, that compares
well with the analytical result $0.628$.
The good agreement of the analytical and Monte Carlo results
indicate that the analytical picture is accurate.
It thus provide a simple and faithful description of device
dynamics.

%
%

\begin{figure}
%
    \centerline{
   \includegraphics[height=7cm,angle=-90]{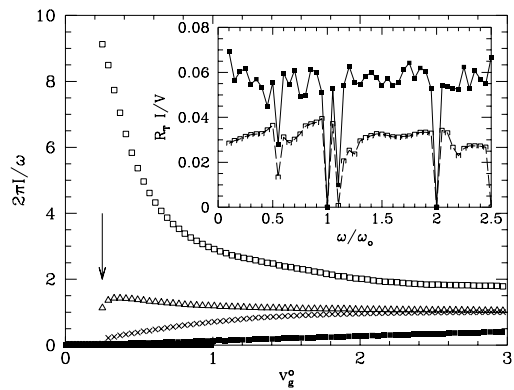}
    }
    \caption{
    DC current response to an oscillating gate voltage
    $v_g=-1/2+v_g^o \sin(\omega t)$ as a function of $v_g^o$.
    $v=1/2$,  $\omega/\omega_o=.1$, (same symbols of \refF{fig3})
    The arrow indicates the threshold to the conducting region.
    Inset: Frequency dependence of the DC current for $v_g^o=1.05$ the threshold value.
    Structures at $\omega/\omega_o=0.5$, 1 and 2 are clearly visible.
    }
\label{fig3}
\end{figure}

%
The presence of the mechanical bistability can be experimentally
probed for $v<1$ by modulating the gate voltage around -1/2:
$v_g=-1/2+v_g^o\, \sin(\omega t)$.
According to the picture given before if $v_g^o<(1-v)/2$, no current
should flow, since the island remains in the same stable or
meta-stable position all the time: $v_g$ varies inside the dashed
diamond of \refF{fig2}.
For $v_g^o > (1-v)/2$ the island is instead always released from the
metastable state, and it is free to oscillate between the two
positions $x=-1$ and $x=0$.
This implies that some current can flow through the device.
The resulting DC current as a function of $v_g^o$ is thus
discontinuous at $v_g^o=(1-v)/2$.
In order to verify this idea quantitatively we resort to
a Monte Carlo simulation for this non stationary situation.
Results are shown in \refF{fig3} for $v=1/2$ and
$\omega/\omega_o=0.1$.
For $v_g^o \gg (1-v)/2=1/4$ we find that the current is one electron
per cycle, regardless of the value of $\Gamma_o$.
The value of the discontinuity depends instead strongly on
$\Gamma_o$.
For $\Gamma_o\gg1$ many electrons can flow during the swing of
the oscillator between the two stable solutions.
In the opposite limit of $\Gamma_o\ll1$, the discontinuity is strongly
reduced.
Nevertheless also in this case the nanomechanical nature of the
current can be probed by studying the DC current as a function of
the external frequency $\omega/\omega_o$.
The inset shows the frequency dependence of the current at the
threshold ($v_g^o=1.05/4$) for large and small $\Gamma_o$.
In both cases dips are present when $\omega/\omega_o=0.5$, 1 or $2$,
indicating the resonant response of the mechanical degree of
freedom.

%
%
%
%

%
%

In conclusion we have shown that the coupling to a mechanical degree
of freedom can lead to a current suppression at the degeneracy point.
A bistability appears for $eV<E_E$ and we have shown how this can be
detected by using an AC gate voltage.
We also presented an analytic theory that allows to calculate with
quantitative accuracy the current in the whole conducting region
($eV>E_E$).
A gap at the degeneracy point has been already observed
experimentally in different nano-mechanical devices
\cite{park:2000,weig:2004}.
The miniaturization of the devices will allow to increase
further the value of $E_E$.
Our predictions can be useful to test if the observed phenomena are
really related to nanomechanical effects.

We acknowledge support from the French Agence Nationale Recherche (project JCJC06\_137869)
and IdNano (Grenoble).

\bibliography{biblioNEMS}

\end{document}